\pdfoutput=1
\documentclass[11pt]{article}
\usepackage[]{EMNLP2023}
\usepackage{times}
\usepackage{latexsym}
\usepackage[T1]{fontenc}
\usepackage[utf8]{inputenc}
\usepackage{microtype}
\usepackage{enumitem}
\usepackage{inconsolata}
\usepackage{pifont} 
\usepackage{graphicx} %
\usepackage{float} %
\usepackage{afterpage}
\usepackage{placeins}
\usepackage{pifont} 
\usepackage{graphicx} 
\usepackage{array} 
\usepackage{threeparttable}
\usepackage{threeparttablex}
\usepackage{booktabs}
\usepackage{threeparttable}
\usepackage{adjustbox}
\usepackage{pifont}
\usepackage{float}
\usepackage[utf8]{inputenc}
\usepackage{newunicodechar} 
\newunicodechar{≥}{$\geq$}
\usepackage{xcolor}
\usepackage{enumitem}
\usepackage{tcolorbox}
\usepackage[framemethod=TikZ]{mdframed}
\usepackage{soul, color}

\title{SouLLMate: An Adaptive LLM-Driven System for Advanced Mental Health Support and Assessment, Based on a Systematic Application Survey}

\author{Qiming Guo$^{*1}$, Jinwen Tang$^{*2}$, Wenbo Sun$^3$, Haoteng Tang$^4$, Yi Shang$^2$, and Wenlu Wang$^1$ \\
$^1$Texas A\&M University - Corpus Christi, Corpus Christi, TX, USA \\
$^2$University of Missouri, Columbia, MO, USA $^3$Delft University of Technology, Delft, Netherlands \\
$^4$University of Texas Rio Grande Valley, Edinburg, TX, USA\\
$^1$ qguo2@islander.tamucc.edu, wenlu.wang@tamucc.edu\\
$^2$ \{jt4cc, shangy\}@umsystem.edu  $^3$ w.sun-2@tudelft.nl  $^4$ haoteng.tang@utrgv.edu \\}

\begin{document}
\maketitle
\begin{abstract}

Mental health issues significantly impact individuals' daily lives, yet many do not receive the help they need even with available online resources. This study aims to provide accessible, stigma-free, personalized, and real-time mental health support through cutting-edge AI technologies. It makes the following contributions: (1) Conducting an extensive survey of recent mental health support methods to identify prevalent functionalities and unmet needs. (2) Introducing SouLLMate, an adaptive LLM-driven system that integrates LLM technologies, Chain, Retrieval-Augmented Generation (RAG), prompt engineering, and domain knowledge. This system offers advanced features such as Suicide Risk Detection and Proactive Guidance Dialogue, and utilizes RAG for personalized profile uploads and Conversational Information Extraction. (3) Developing novel evaluation approaches to assess preliminary assessments and suicide risk detection, utilizing annotated real-life interview data and professionally labeled datasets indicating suicide tendencies. (4) Proposing Key Indicator Summarization (KIS) and Proactive Questioning Strategy (PQS) methods to enhance model performance and usability through context-sensitive response adjustments and semantic coherence evaluations. This study contributes to advancing mental health support technologies, potentially improving the accessibility and effectiveness of mental health care globally.
\end{abstract}

\section{Introduction}

According to the National Institute of Mental Health (NIMH), nearly 20\% of adults experience a mental illness each year \cite{NIMH_Mental_Illness}, with the availability of mental health care and support remaining inadequate \cite{bland1997help}. Disorders like depression and anxiety significantly affect individuals' lives, disrupting their ability to work, maintain relationships, and participate in their communities. \textbf{To address these significant gaps in mental health support,} we introduce our SouLLMate system 
~\footnote{*The first two authors contributed equally to this work.}
\footnote{Note: The detailed project code and prompts for SouLLMate are at GitHub: \href{https://github.com/QM378/SouLLMate}{https://github.com/QM378/SouLLMate}; Also a brief video introducing SouLLMate can be found at: \newline
\href{https://www.youtube.com/watch?v=crxj0VjyFAE}{https://www.youtube.com/watch?v=crxj0VjyFAE}}
, which utilizes large language models (LLMs) to enable real-time, supportive communication, making services more accessible, personalized, and private.

\begin{table*}
\vspace{-4mm}
\begin{ThreePartTable}
\resizebox{\textwidth}{!}{%
\setlength{\tabcolsep}{2pt}
\renewcommand{\arraystretch}{0.85} 
\scriptsize

\begin{tabular}{l|cccccccccc}
\hline
Functionalities$\textbackslash$Apps & \textbf{GMH-QA} & \textbf{PMH-A} & \textbf{CIE} & \textbf{SRD} & \textbf{PGD} & \textbf{PMHR} & \textbf{RAG} & \textbf{Price} & \textbf{Methods} \\ \hline
Happify \cite{happify2022} & \scalebox{0.75}{\ding{51}} & & \scalebox{0.75}{\ding{51}} & & & \scalebox{0.75}{\ding{51}} & & \$139.99/Y & IBM technology \\
Headspace \cite{headspace2024} & \scalebox{0.75}{\ding{51}} & & \scalebox{0.75}{\ding{51}} & & & & & \$69.99/Y & N/A \\
Calm \cite{calm2024} & \scalebox{0.75}{\ding{51}} & & & & & & & \$69.99/Y & Q bot \\
MindShift \cite{mindshift2024} & \scalebox{0.75}{\ding{51}} & & & & & & & F+D & GPT 3.5 \\
Shine \cite{shineapp2024} & \scalebox{0.75}{\ding{51}} & & & & & & & IAP & N/A \\
Bearable \cite{bearable} & \scalebox{0.75}{\ding{51}} & & & & & \scalebox{0.75}{\ding{51}} & & IAP & N/A \\
Reflectly \cite{reflectly} & & & & & & \scalebox{0.75}{\ding{51}} & & IAP & N/A \\
Rise Up \cite{riseup} & \scalebox{0.75}{\ding{51}} & & & & & & & IAP & N/A \\
Rootd \cite{rootd} & \scalebox{0.75}{\ding{51}} & & & & & & & IAP & N/A \\
Sanvello \cite{sanvello2024} & \scalebox{0.75}{\ding{51}} & & & & & \scalebox{0.75}{\ding{51}} & & \$199.99/Y & N/A \\
MoodTools \cite{moodtools2024} & \scalebox{0.75}{\ding{51}} & \scalebox{0.75}{\ding{51}} & & & & & & \$139.99/Y & N/A \\
Breathe2Relax \cite{breathe2relax} & & & \scalebox{0.75}{\ding{51}} & & & & & Free & N/A \\
Panic Attack \cite{panicattack2024} & \scalebox{0.75}{\ding{51}} & & & & & & & IAP & N/A \\
Brain.fm \cite{brainfm2024} & & & \scalebox{0.75}{\ding{51}} & & & & & \$9.99/M & N/A \\
Ada \cite{ada2024} & \scalebox{0.75}{\ding{51}} & & & & & & & IAP & N/A \\
Amaha \cite{amaha2024} & \scalebox{0.75}{\ding{51}} & & \scalebox{0.75}{\ding{51}} & & & & & IAP & AI/ML \\
Replika \cite{replika2024} & \scalebox{0.75}{\ding{51}} & & \scalebox{0.75}{\ding{51}} & & & & & IAP & AI/ML/LLM \\
Serenity \cite{serenity2024} & \scalebox{0.75}{\ding{51}} & & & & & \scalebox{0.75}{\ding{51}} & \scalebox{0.75}{\ding{51}} & F & GPT 3.5 \\
Woebot \cite{woebot2024} & \scalebox{0.75}{\ding{51}} & & \scalebox{0.75}{\ding{51}} & & & & & IAP & Gen AI \\
Wysa \cite{wysa2024} & \scalebox{0.75}{\ding{51}} & \scalebox{0.75}{\ding{51}} & \scalebox{0.75}{\ding{51}} & & & & & IAP & LM \\
Youper \cite{youper2024} & \scalebox{0.75}{\ding{51}} & \scalebox{0.75}{\ding{51}} & \scalebox{0.75}{\ding{51}} & & & \scalebox{0.75}{\ding{51}} & & IAP & LLM +DL \\
MindDoc \cite{minddoc2024} & \scalebox{0.75}{\ding{51}} & \scalebox{0.75}{\ding{51}} & & & & \scalebox{0.75}{\ding{51}} & & IAP & N/A \\
Dr. Sila \cite{drsila2024} & \scalebox{0.75}{\ding{51}} & \scalebox{0.75}{\ding{51}} & & & & & & IAP & N/A \\
Pocketcoach \cite{pocketcoach2024} & \scalebox{0.75}{\ding{51}} & & & & & & & IAP & N/A \\
\hline
\textbf{SouLLMate}(Our Method) & \scalebox{0.75}{\ding{51}} & \scalebox{0.75}{\ding{51}} & \scalebox{0.75}{\ding{51}} & \scalebox{0.75}{\ding{51}} & \scalebox{0.75}{\ding{51}} & \scalebox{0.75}{\ding{51}} & \scalebox{0.75}{\ding{51}} & Local LLM/API usage & LLMs \\
\hline
\end{tabular}%
}

\caption{Functionality comparison between various mental health apps.}
\label{table:table1}
\begin{tablenotes}
\scriptsize
\item 
\textbf{GMH-QA} (General Mental Health QA): Common mental health questions and answers.
\textbf{PMH-A} (Preliminary Mental Health Assessment): Initial evaluation of user's mental health.
\textbf{CIE} (Conversational Information Extraction): User information gathering via chat.
\textbf{SRD} (Suicide Risk Detection): Immediate crisis intervention and risk assessment.
\textbf{PGD} (Proactive Guidance Dialogue): Active engagement and guidance based on user input.
\textbf{PMHR} (Personal Mental Health Reporting): Mental health report creation from user data.
\textbf{RAG} (Retrieval-Augmented Generation): Enhanced information retrieval and generation based on user-provided documents and queries.
Price Terms:
\textbf{F} (Free): No cost to use.
\textbf{IAP} (In-App Purchases): Free to download with optional paid features or content.
\textbf{F+D} (Free with Donations): Free to use with optional donations accepted.
*Human-delivered services are excluded for fair comparisons.
\end{tablenotes}
\end{ThreePartTable}
\vspace{-4mm}
\end{table*}

\textbf{Effective communication is crucial in mental health care but often falls short of meeting patient needs.} Research has shown that effective communication can significantly boost well-being and help manage mental health conditions like depression and anxiety \cite{galderisi2017proposed, kiesler1979interpersonal}. It serves as a vital support, enabling individuals to navigate life's challenges and share essential resources effectively. However, falling into negative communication patterns can exacerbate conditions like depression and eating disorders \cite{chiariello1995patterns, prescott2002eating}. Furthermore, many people encounter barriers such as stigma and high treatment costs, which prevent them from seeking help  \cite{bland1997help}. These challenges underscore the urgent need for better communication strategies that can effectively support mental health, facilitate resource sharing, and help reduce the stigma associated with seeking help.

\textbf{Recent innovations have transformed how we manage mental health, but challenges remain in delivering effective support.} Platforms like social media, mobile apps, and chatbots now provide real-time, low-cost options for individuals to share their experiences and manage mental health conditions in a judgment-free environment \cite{naslund2020social, neary2018state, abd2021perceptions}. While these tools greatly enhance accessibility and reduce stigma, they often face challenges in delivering reliable content and supportive human-like communication in real-time.

To address these challenges, we introduce SouLLMate, a system built on off-the-shelf large language models to improve mental health support and can easily adapt to different scenarios. SouLLMate offers real-time, human-like interactions and comes equipped with tools to assess and support a variety of mental health conditions, such as depression, anxiety, PTSD, and stress-related disorders.
Specifically, our contributions are as follows:

\begin{itemize}[itemsep=-1.5pt,parsep=2pt]
    \item We conducted an extensive survey of recent methods supporting mental health, revealing common functionalities among existing approaches as well as important needs that these methods had yet to fulfill compared to the assistance obtainable from human experts. This can aid future researchers in designing more comprehensive methods.
    
    \item We introduced \textbf{SouLLMate}, an adaptive LLM-driven system for advanced mental health support and assessment. This system integrates LLM technologies, LangChain, Retrieval-Augmented Generation (RAG), prompt engineering, and domain expertise knowledge in mental health. SouLLMate offers functionalities beyond General Mental Health Question-Answer (QA) and Preliminary Mental Health Assessment, including Suicide Risk Detection and Proactive Guidance Dialogue. Unlike existing methods, our system employs RAG to allow users to upload personal profiles and utilizes Conversational Information Extraction to collect key linguistic indicators, enhancing various support functions, such as Personal Mental Health Reporting.

    \item Given the scarcity of evaluation methods for mental health support systems, we developed novel approaches to assess the Preliminary Mental Health Assessment and Suicide Risk Detection functionalities. Our evaluation methodology utilizes annotated real-life interview data and professionally labeled suicide tendency datasets to rigorously assess these critical functions.
    
    \item We proposed two novel methods to enhance model performance and usability: 1) KIS: a context-sensitive adjustment mechanism that modifies responses based on user history and current session data, and 2) PQS: a semantic coherence evaluation strategy ensuring logical conversation flow and sustained user engagement over extended interactions.
    
\end{itemize}

The remainder of this paper is organized as follows: Section 2 investigates existing methods; Section 3 introduces the design of the SouLLMate system; Section 4 evaluates our system; and Section 5 concludes the paper.

\section{Investigation And Background}

Language models are crucial tools for supporting research in the field of mental health. The rapid development of language models \cite{gpt4openai, metallama3} in the past two years has led to significant innovations and breakthroughs across numerous domains. 
We summarize our findings in Table \ref{table:table1}. We categorize the available applications based on supported services, functionalities, and prices. 
As illustrated in Table \ref{table:table1}, our investigation reveals that most methods incorporate General Mental Health Question-Answering (GMH-QA) functionality. Approximately half of the surveyed methods also include Preliminary Mental Health Assessment (PMH-A), Conversational Information Extraction (CIE), and Personal Mental Health Reporting (PMHR) features. Additionally, a detailed discussion regarding how these listed methods align with the broad needs of mental health, as identified in prior literature, is provided below \cite{martin2011effectiveness, van2016validation, rothi2006mental, lindow2020youth, thornicroft2016evidence}.

\textbf{Real-Time Communication} is essential for providing immediate support. Apps like Happify utilize a chatbot named Taylor for continuous mental health monitoring and personalized assistance \cite{happify2022}, while Headspace and Wysa enhance user engagement with virtual assistants Frankie and an AI conversational interface, respectively \cite{headspace2024, wysa2024}. \textbf{Preliminary Assessments} play a critical role in mental health apps; platforms such as Wysa, Youper, and MindShift use AI to implement recognized assessment tools like the GAD-7 and PHQ-9, facilitating effective symptom tracking and progress monitoring \cite{wysa2024, youper2024, mindshift2024}. \textbf{Educational Content and Resources} are prominently featured in apps like Happify and Headspace, which provide scientifically-backed exercises and content \cite{happify2022, headspace2024}, and Calm, which offers in-depth insights into mental health care plans \cite{calm2024}. \textbf{Encouraging Help-Seeking Behavior}, Reflectly, Shine, and MindShift actively engage users with daily reflections and provide resources to effectively manage coaching sessions \cite{reflectly, shineapp2024, mindshift2024}. Lastly, \textbf{Reducing Stigma} is a key focus for apps like MindShift, which offers anonymous interactions, and Wysa and Bearable, which prioritize privacy and secure communications to foster open discussions and effective symptom management without fear of judgment \cite{mindshift2024, wysa2024, bearable}.

\textbf{Overall, our investigation reveals that, despite some apps performing well in certain aspects, current systems exhibit significant limitations:}

\begin{itemize}[itemsep=-3pt,parsep=0pt]
    \item Unlike human experts who can leverage personal profiles for tailored understanding, most systems lack effective Retrieval-Augmented Generation (RAG) capabilities, diminishing their efficacy and utility.
    \item Human experts initiate proactive dialogue based on individual interests, fostering open communication. In contrast, most existing systems offer only basic, passive question-answering functionalities, often failing to address deep-seated needs.
    \item Human experts can identify and intervene in cases of suicidal tendencies during real-world interactions. However, existing systems largely lack functionalities for detecting such critical conditions.
    \item Current systems and methods lack reliable, evidence-based approaches to validate their effectiveness \cite{nicholas2015mobile, van2015quality}.
\end{itemize}

\FloatBarrier
\begin{figure*}[!htbp]
  \centering
  \includegraphics[width=1 \linewidth]{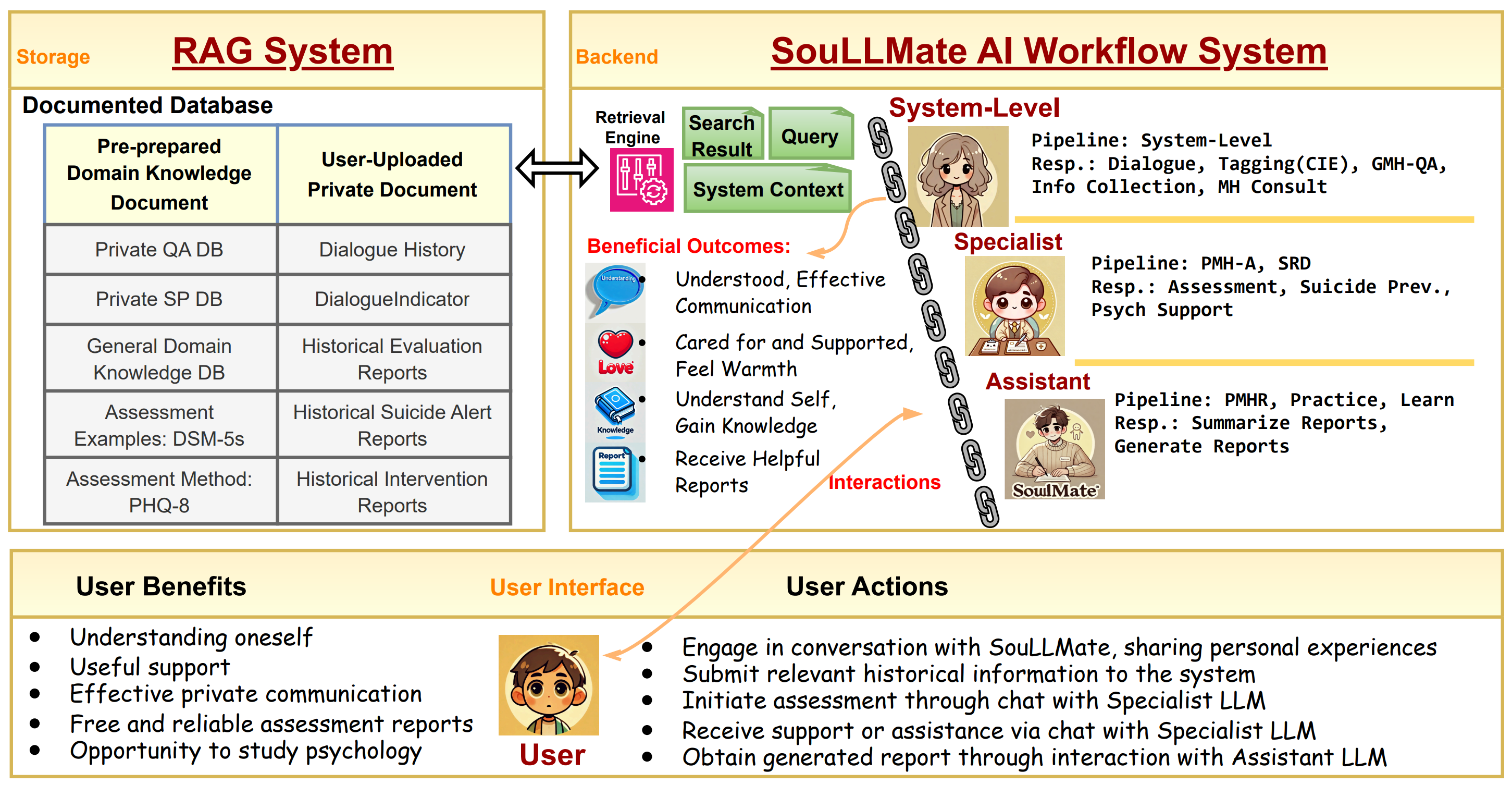}
  \caption{Structure of SouLLMate AI System.}
  \label{fig:workflow}
\end{figure*}

\textbf{Fortunately, advanced large language models (LLMs) show promising potential for enhancing mental health support.} Recent LLMs such as GPT-4 \cite{gpt4openai}, Llama 3 \cite{metallama3}, and Claude 3.5 \cite{claude35sonnet} have demonstrated impressive performance in complex reasoning tasks, achieving remarkable results on benchmarks like ARC Challenge and GPQA using zero-shot and Chain of Thought (CoT) methods.

These state-of-the-art LLMs have the potential to transform future health technologies, potentially engaging in complex tasks such as psychological counseling and preliminary screening. \citet{james2023towards} explored the use of GPT-3 to enable mental health professionals and patients with severe mental illnesses to more effectively formulate personalized and measurable treatment goals. \citet{tangshang2024advancing} advanced mental health pre-screening of interview records using a customized GPT-4 model,``Psycho Analyst," integrating DSM-5 and PHQ-8 into a sophisticated binary classification and multi-stage assessment framework, demonstrating high accuracy and broad potential applications.

Our investigation indicates an imminent, significant transformation in the field of mental health support through the application of large language models.

\section{System Design}

\textbf{SouLLMate, designed to deliver assistance that closely mirrors the support provided by human experts}, is capable of (1) Preliminary Assessment (PMH-A), (2) obtaining key information through dialogue like humans (CIE), and (3) understanding a person's historical situation or acquiring more document information through reading documents (RAG). (4) Conducting a Proactive Guidance Dialogue (PGD) for individuals who have a negative desire to talk. (5) Suicide Risk Detection is also one of the functionalities the system aims to implement. (6) Ultimately, it will generate an appropriate report.

\textbf{To implement these functionalities, we utilize}: Large Language Models (LLMs), LangChain, Retrieval-Augmented Generation (RAG), prompt engineering, and domain expertise. As illustrated in Figure \ref{fig:workflow}, the system is structured into three LLM-based levels:
 
\begin{itemize}[itemsep=-2pt,parsep=2pt]
    \item \textbf{System Level}: Implements direct user dialogue, General Mental Health QA (GMH-QA), Conversational Information Extraction (CIE), and RAG functionalities.
    
    \item \textbf{Specialist Level}: Executes Proactive Guidance Dialogue (PGD), Preliminary Mental Health Assessment (PMH-A), and Suicide Risk Detection.
    
    \item \textbf{Assistant Level}: Generates the final report.
\end{itemize}

The following sections will detail the design of the LLM-based workflows.~\footnote{Due to space constraints, the system deployment and usage examples will be presented in the appendix. The detailed prompts used for this system will be available on GitHub.}

\subsection{Design of System Level Workflow}

The System Level workflow is designed as a cyclical dialogue chain. It serves as the initial direct contact enabling the SouLLMate system to connect with the user, after which other components can operate accordingly.

\textbf{\textbullet} Responsibility: The workflow is tasked with engaging in dialogue with users; identifying and extracting indicators and information through conversation; connecting to the database which stores pre-installed knowledge-based documents along with users' historical information and customized documents; and conducting General Mental Health QA (GMH-QA) by invoking RAG to access and utilize database information.

\textbf{\textbullet} Prompt Design: The prompts primarily define the LLM's role as a system-level LLM, outlining its methods and styles of dialogue with users, authority in responses, and the rights to invoke RAG and access or update the user's state.

\textbf{\textbullet} RAG Design: The moment a user interacts with the system, the RAG system is invoked at the System Level to retrieve user history and directly contribute to the dialogue and consultation.

\subsection{Design of Specialist Level Workflow}

The specialist workflow consists of three components, which implement PGD, PMH-A, and SRD functionalities based on the information obtained from the System Level LLM.

\subsubsection{Proactive Guidance Dialogue (PGD)}

\hspace{\parindent} \textbf{\textbullet} Responsibility: PGD proactively guides dialogue, helps users alleviate emotions, and encourages openness. It utilizes RAG functionality to access the database (The database contains knowledge documents, the user’s most recent emotional state, historical assessment data, and past suicidal tendencies).

\textbf{\textbullet} Prompt Design: The prompt primarily defines the LLM's role as an expert, patiently discovering users' mental health needs and actively guiding topics to encourage users to express themselves and open up.

\noindent \subsubsection{Preliminary Mental Health Assessment (PMH-A)}

\hspace{\parindent} \textbf{\textbullet} Responsibility: PMH-A evaluates user-submitted assessment documents or conducts dialogue-based assessments in the current conversation window. Based on the defined assessment prompts, utilize the Patient Health Questionnaire–8 (PHQ-8) \cite{Kroenke2002} to assess and score the obtained text, identifying scores ≥10 as indicative of mental health issues.

\textbf{\textbullet} Prompt Design: The prompt primarily defines the LLM's role as an assessment expert, the questioning strategy in user dialogues, the PHQ-8 scoring criteria for evaluations, and the format for producing assessment outcomes.

\subsubsection{Suicide Risk Detection (SRD)}

\hspace{\parindent} \textbf{\textbullet} Responsibility: This functionality assesses users' suicidal tendencies using historical data, documents in the database, user dialogues, and uploaded information accessed through RAG. It also evaluates these tendencies according to three professional dimensions: intention, behavior, and whether the individual has received support from others. 

\textbf{\textbullet} Prompt Design: The prompt primarily defines the LLM's role, establishes the criteria and basis for utilizing suicidal tendency detection through stages such as intent, ideation, behavior, and plan.

\subsection{Design of Assistant Level Workflow}

\hspace{\parindent} \textbf{\textbullet} Responsibility: Create personalized reports for users to download and keep as records, or they can submit these to mental health professionals to provide pre-screening information and historical records.

\textbf{\textbullet} Prompt Design: The prompt primarily defines the LLM's role as a summarizer and advisor, tasked with generating a comprehensive report based on users' recent assessment results, dialogue history, and current status.

\subsection{Techniques to Enhance Performance}

Here we introduce two methods to enhance the performance (in terms of inference time and accuracy) of mental health pre-screening.

\textbf{\textbullet} \textbf{Key Indicator Summarization (KIS)}: In mental health contexts, which often require long-form dialogues, extended inputs can compromise accuracy and prolong inference times, which cannot be handled by relatively small-size open-source LLM models \cite{sun2024cebench}. To address this, we propose a method that combines prompt engineering with domain expertise to distill key information from historical dialogues. This information is then transformed into a concise list of phrases, forming a new dataset incorporated into the assessment pipeline. By focusing on these phrases instead of entire conversations, this approach enables assessments for mental disorders and suicide risk with reduced computational demands. Consequently, it alleviates storage and hardware limitations, allowing deployment on relatively small-size open-source models. 

Using the local LLaMA       3 8B model, SouLLMate processed 187 interview datasets in a single session with a valid response rate of 74.3\%. After incorporating the KIS method, the valid response rate increased to 100\%(as shown in Table \ref{tab:example}) , aligning with the performance of Mixtral8*7B and GPT.

\textbf{\textbullet} \textbf{Proactive Questioning Strategy (PQS)}: Users often narrate their experiences during conversations, which, while relieving stress and emotions, may not be highly beneficial for the evaluation model. We proposed a proactive questioning strategy, mimicking the inquisitive assessing approach used by psychologists. By pre-defining the LLM's role and behavior in the prompt, the pre-screening process can be expedited, and more relevant information can be gathered. In the SouLLMate system, we designed the PQS (Proactive Questioning Strategy) functionality. Examples of how it works are shown in Table~\ref{tab:example}.

\begin{table}[h!]
  \centering
  \begin{minipage}{\columnwidth}
    \centering
    \setlength{\tabcolsep}{3pt}
    \renewcommand{\arraystretch}{1.2} 
    \small
    \begin{tabular}{m{0.45\columnwidth}|m{0.45\columnwidth}}
      \hline
      \multicolumn{2}{l}{\textbf{User Input:} Today was just one of those days...} \\
      \hline
      Original Response & Response with PQS  \\
      \hline
      Hope tomorrow is better for you! & It sounds like today was tough. Want to talk about what happened? \\
      \hline
    \end{tabular}
    \caption{Response Examples}
    \label{tab:example}
    \vspace{-4mm}
  \end{minipage}
\end{table}

Those techniques allow for more efficient extraction of key information from lengthy dialogues, thereby assisting in effective evaluation and pre-screening.

\section{Evaluation of the System}
We utilize a range of off-the-shelf large language models to conduct a function-oriented  assessment, based on two annotated datasets (detailed below). 

\subsection{Depression Assessment Function}
The DAIC-WOZ database, designed to assist in studying psychological distress such as anxiety, depression, and PTSD, was utilized to evaluate the performance of our SouLLMate system in assessing depression. The dataset comprises interviews conducted by a virtual interviewer controlled by a human, and includes a total of 187 labeled interviews \cite{gratch2014distress}. The evaluation was performed on the testing set with 47 data entries. The results are presented in Table~\ref{tab:pre_diagnosis_accuracy}.
\begin{table}[h]
  \centering
  \small
  \begin{tabular}{lccc}
    \hline
    \textbf{Model} & \textbf{Valid (\%)} & \textbf{MAE} & \textbf{RMSE} \\
    \hline
    LLaMA 3 8B    & 74 & 4.21 & 4.98 \\
    LLaMA 3 8B + KIS      & 100 & 4.14 & 4.40 \\
    Mixtral 8x7B & 100 & 3.93 & 5.05 \\
    GPT-4    & 100 & 3.28 & 4.39 \\
    \hline
  \end{tabular}
  \caption{Performance of Pre-Screening Assessment}
  \label{tab:pre_diagnosis_accuracy}
  \vspace{-4mm}
\end{table}

\subsection{Suicide Tendency Assessment Function}
The Reddit dataset represents a characterization of Time-variant and Time-invariant Assessments of Suicidality on Reddit using the Columbia Suicide Severity Rating Scale (C-SSRS) \cite{gaur2021characterization}. We evaluate our system's performance in detecting indicators of suicidal intentions and behaviors through text analysis. This dataset includes 448 Reddit users and utilizes the Columbia Suicide Severity Rating Scale \cite{posner2011columbia} for annotations. For the evaluation, 10\% of the data was randomly selected for multi-category testing, which includes categories for Ideation, Behavior, and Supportive actions (with original dataset attempts reclassified as Behavior). The results of this evaluation are presented in Table~\ref{tab:suicide_tendency_accuracy}.

\begin{table}[h]
  \centering
  \setlength{\tabcolsep}{3pt}
  \small
  \begin{tabular}{lccc}
    \hline
    \textbf{Model} & \textbf{Macro Precision} & \textbf{Accuracy} & \textbf{Macro F1} \\
    \hline
    GPT-3.5      & 0.48 & 0.58 & 0.53 \\
    GPT-4o      & 0.83 & 0.67 & 0.68 \\
    GPT-4-turbo      & 0.82 & 0.64 & 0.63 \\
    \hline
  \end{tabular}
  \caption{Performance of Suicide Tendency Detection}
  \label{tab:suicide_tendency_accuracy}
\vspace{-4mm}
\end{table}

\section{Conclusion}

Our study is among the few that focus on utilizing large language models to enhance mental health support. Through investigating existing methodologies and developing the adaptive SouLLMate system, we have established a comprehensive pipeline offering robust consultations, preliminary assessments, and suicide interventions. The SouLLMate system empowers users to either delete their assessment data or retain it for personal use, addressing long-standing challenges such as high costs, insufficient healthcare resources, and privacy concerns.

This system not only charts a new course for mental health support but also serves as a blueprint for the future integration of AI in healthcare. We believe that advancements in AI will render healthcare resources more affordable, accessible, secure, private, and unbiased, making them readily available to those in need across various platforms. Additionally, the development of more advanced models, such as video large models and multimodal large models, will significantly enhance the usability of mental health support tools like SouLLMate, democratizing the benefits of research in this field for everyone.

\section*{Limitations}

\textbf{Model Constraints:} Effective mental health pre-screening requires models with high contextual understanding and inference capabilities. However, limitations in model selection restrict the research scope. \textbf{Data Scarcity:} The scarcity of publicly available, professionally annotated data hampers the development and validation of mental health models. The available data is often sparse, unannotated, or excessively lengthy, posing significant challenges. \textbf{Resource Costs:} Mental health research is resource-intensive. Using advanced models like the Mixtral 8*7B on high-end servers can take up to an hour for a single inference, limiting prompt testing and workflow adjustments due to high operational costs.

\section*{Ethics Statement}

Our research does not utilize personal data, relying solely on open-source data and methods, thus avoiding any infringement on personal privacy. When using our system, users have the option to upload or withhold their personal data from the model. Additionally, they can delete the DB file to erase their usage history, thereby protecting their personal privacy.

This system employs various technologies to support mental health, functioning as a supportive soul companion. However, it is important to note that it does not possess the same authority or capabilities as licensed professionals. The system should not be used as a substitute for clinical interventions or professional medical advice.

\section*{Acknowledgements}

This work was partially supported by the National Science Foundation (NSF) under Award No. 2112631, and by the US Department of Transportation (USDOT) Tier-1 University Transportation Center (UTC) Transportation Cybersecurity Center for Advanced Research and Education (CYBER-CARE). (Grant No. 69A3552348332).

\bibliography{emnlp2023}
\bibliographystyle{acl_natbib}

\appendix

\clearpage 
\section{Appendix: Deployment and Usage Example}

\subsection{System Deployment}

In this demo, we utilized the simple Panel library to showcase our system. Our demo environment includes Windows 11, Python 3.12, and the EDGE browser, primarily using the Panel library to build the interactive interface that displays the system. Nevertheless, by adjusting the UI and running platform, it has the potential to: \textbf{1)} operate on various platforms and \textbf{2)} demonstrate flexibility that could enable the system to meet diverse needs and consider cultural diversity in its interactions.

\subsection{Use Case for Preliminary Mental Health Assessment}

\label{subsec:Appdendix A.2}

\begin{mdframed}[roundcorner=10pt,backgroundcolor=yellow!5!white!95, linecolor=pink!80!white, middlelinewidth=2pt, innerleftmargin=10pt, innerrightmargin=10pt, innertopmargin=10pt, innerbottommargin=10pt, font=\fontsize{8pt}{10pt}\selectfont]
\textbf{1. Total score:} 20

\textbf{2. Individual scores:}
\begin{enumerate}[itemsep=0pt, parsep=0pt, topsep=0pt, partopsep=0pt]
    \item Lack of interest in activities: 3;
    \item Feelings of depression or hopelessness: 3;
    \item Sleep issues: 3;
    \item Low energy: 3;
    \item Changes in appetite: 1;
    \item Negative self-perception: 2;
    \item Concentration difficulties: 2;
    \item Unusual movement or speech patterns: 3;
\end{enumerate}

\textbf{3. Explanation:}
The individual's diary entry indicates significant challenges with their mental health over the past two weeks. They express a lack of interest in activities they used to enjoy, feelings of depression and hopelessness, sleep issues interfering with their rest, low energy levels, some changes in appetite, negative self-perception, concentration difficulties, and potential signs of anxiety with unusual movement or speech patterns. These scores collectively suggest a high level of distress and potential need for professional support to address these concerns and improve their well-being.
\end{mdframed}

\captionof{figure}{One Use Case for Preliminary Mental Health Assessment} 
\label{subsec:figure a1}

Figure \ref{subsec:figure a1} shows the Preliminary Mental Health Assessment results based on the PHQ-8 protocol. It displays total and item scores with interpretation, demonstrating efficiency and accuracy. This comprehensive output facilitates quick, informed decisions on mental health status.

\FloatBarrier
\begin{figure*}[tp] 
  \centering
  \includegraphics[width=0.95\linewidth]{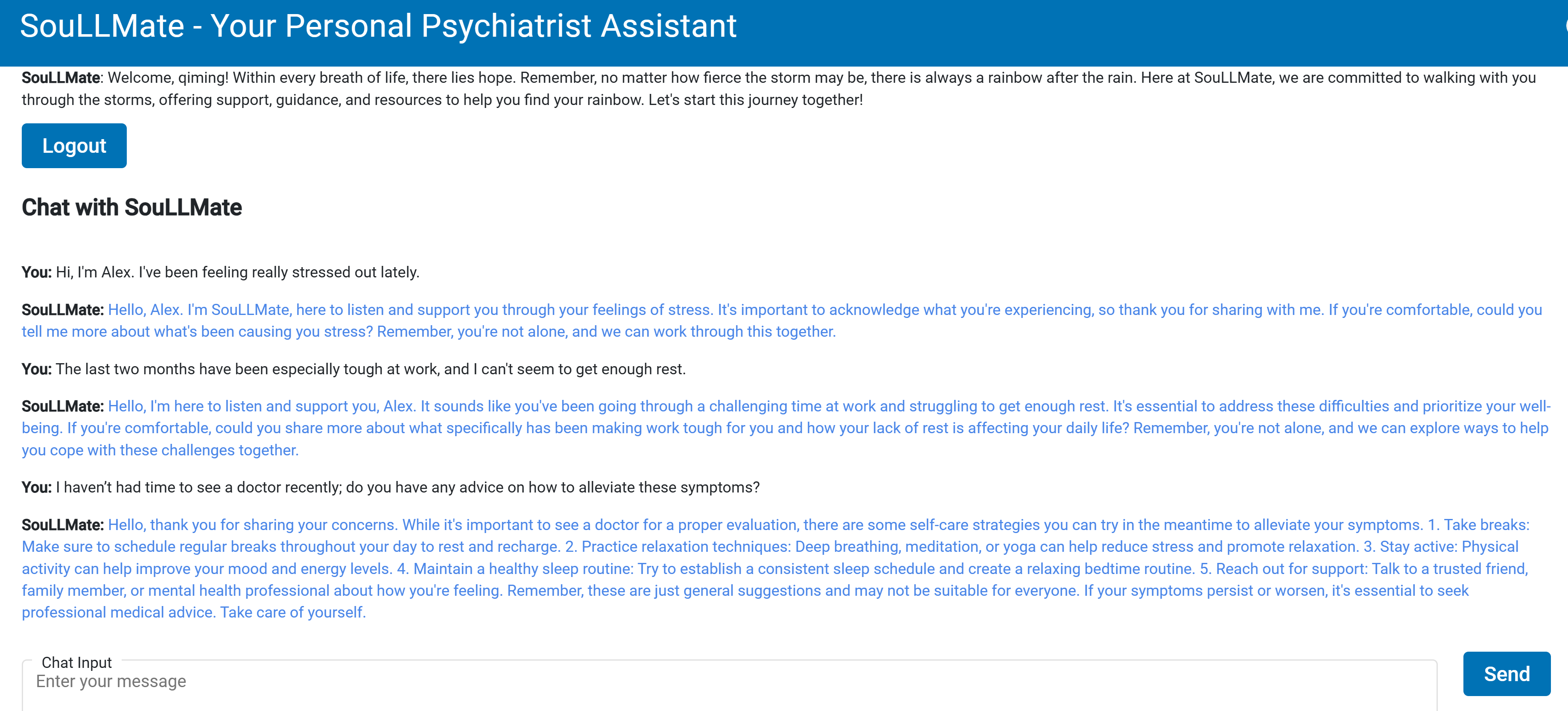}
  \caption{Use Case for General Mental Health QA.}
  \label{fig:Use Case for General Mental Health QA.}
\end{figure*}

\subsection{Use Case for Proactive Guidance Dialogue}

SouLLMate's responses, as shown in Figure \ref{figure:One Use Case for Proactive Guidance Dialogue}, demonstrate empathy, insight, and practical advice. Key aspects include:

\begin{mdframed}[roundcorner=10pt,backgroundcolor=yellow!5!white!95, linecolor=pink!80!white, middlelinewidth=2pt, innerleftmargin=10pt, innerrightmargin=10pt, innertopmargin=10pt, innerbottommargin=10pt, font=\fontsize{8pt}{10pt}\selectfont]

\textcolor{blue}{You:} I guess I'm here because I sometimes feel a bit overwhelmed. It's not all the time, just now and then.

\textcolor{orange}{SouLLMate:} Hello, it's completely normal to feel overwhelmed at times, especially with everything going on in our lives. \textcolor{purple}{It's great that you're aware of these feelings and willing to talk about them. Remember, it's okay to reach out for help when you need it, whether that's talking to someone you trust or seeking professional support. You're not alone in this, and we can work together to find healthy ways to manage those overwhelming moments.}[\textsuperscript{[1]}] Take care of yourself.

\textcolor{blue}{You:} Lately, I find it harder to enjoy things I used to like; not sure why that's happening.

\textcolor{orange}{SouLLMate:} Hello, it's understandable to feel this way when you notice a decrease in your enjoyment of activities you used to like. \textcolor{purple}{Changes in our interests can be influenced by various factors, including stress, mood, or life circumstances. It's great that you're reflecting on these feelings and seeking insight into why it might be happening.}[\textsuperscript{[2]}] Remember, it's okay to explore these emotions further and consider reaching out for support if needed. You're not alone, and we can work together to understand and address these changes. Take care of yourself.

\textcolor{blue}{You:} At night, I often find myself thinking about things I didn't finish during the day. It keeps me up.

\textcolor{orange}{SouLLMate:} Hello, it's common to experience racing thoughts at night, especially when reflecting on unfinished tasks from the day. \textcolor{purple}{This can make it challenging to relax and fall asleep. One helpful strategy is to create a bedtime routine that includes wind-down activities like reading or gentle stretching to signal to your body that it's time to rest. Writing a to-do list before bed can also help clear your mind of lingering thoughts.}[\textsuperscript{[3]}] Remember, it's essential to prioritize your rest for overall well-being. If these thoughts persist, consider speaking to a mental health professional for additional support. Take care of yourself.

\textcolor{blue}{You:} Sometimes, I wonder if it’s normal to feel this way or if I should be doing something about it.

\textcolor{orange}{SouLLMate:} Hello, it's completely normal to have these thoughts and feelings, especially when you're experiencing challenges with your mental health. \textcolor{purple}{It's essential to acknowledge what you're going through and consider seeking help if you feel it's impacting your well-being. You're taking a positive step by reflecting on these concerns, and it shows strength to consider ways to address them.}[\textsuperscript{[4]}] Remember, it's okay to prioritize your mental health and explore support options that can help you navigate these emotions. You're not alone, and we can work together to find ways to improve your well-being. Take care of yourself.

\end{mdframed}

\captionof{figure}{One Use Case for Proactive Guidance Dialogue} 
\label{figure:One Use Case for Proactive Guidance Dialogue}

\begin{itemize}[itemsep=-2pt,parsep=2pt]

\item [\textcolor{purple}{1}] \textbf{Empathetic Validation:} Normalizing emotions and encouraging disclosure.
\item [\textcolor{purple}{2}] \textbf{Cognitive Exploration:} Guiding self-reflection on emotional states.
\item [\textcolor{purple}{3}] \textbf{Behavioral Intervention:} Offering strategies for stress-induced insomnia.
\item [\textcolor{purple}{4}] \textbf{Psychoeducation:} Promoting proactive mental health management.
\end{itemize}

These interactions illustrate SouLLMate's ability to provide comprehensive psychological support through evidence-informed dialogues.

\subsection{Use Case for Suicide Risk Detection}

As shown in Figure \ref{figure:Use Case for Suicide Risk Detection}, this assessment effectively evaluates suicide risk by quantifying key factors such as intent, ideation, and behavior. It accurately identifies the absence of suicidal tendencies while recognizing the supportive nature of the response, demonstrating its efficacy in mental health dialogues.

\begin{mdframed}[roundcorner=10pt,backgroundcolor=yellow!5!white!95, linecolor=pink!80!white, middlelinewidth=2pt, innerleftmargin=10pt, innerrightmargin=10pt, innertopmargin=10pt, innerbottommargin=10pt, font=\fontsize{8pt}{10pt}\selectfont]
\textbf{Content:} "It's not a viable option, and you'll be leaving your wife behind. You'd pain her beyond comprehension. It sucks worrying about money, I know that firsthand. It can definitely feel hopeless, as you seem to be tiredly aware of. Your wife might need to chip in financially. I know time is an issue, but even 10-15 hours a Asthenia could alleviate a lot of the pressure. In the meantime, get your shit together - write that resume tomorrow. No excuses, get it done and send it out. Whether you believe in some sort of powerful being or force governing things or not, things really do work themselves out. This is a big test for you, and youll pull through. Just try to stay as positive as you can and everything will work out."

\textbf{Generated Results:}
\begin{itemize}[itemsep=-2pt,parsep=2pt]

    \item Suicide intent: 0
    \item Suicide phrase: NA
    \item Passive Ideation: 0
    \item Active Ideation: 0
    \item Intent: 0
    \item Plan: 0
    \item Behavior: 0
    \item Supportive: 1
    \item User: user-0
\end{itemize}
\end{mdframed}

\captionof{figure}{Use Case for Suicide Risk Detection} 
\label{figure:Use Case for Suicide Risk Detection}

\subsection{Use Case for Report Generation}

\FloatBarrier
\begin{figure}[H] 
  \centering
  \includegraphics[width=0.95\linewidth]{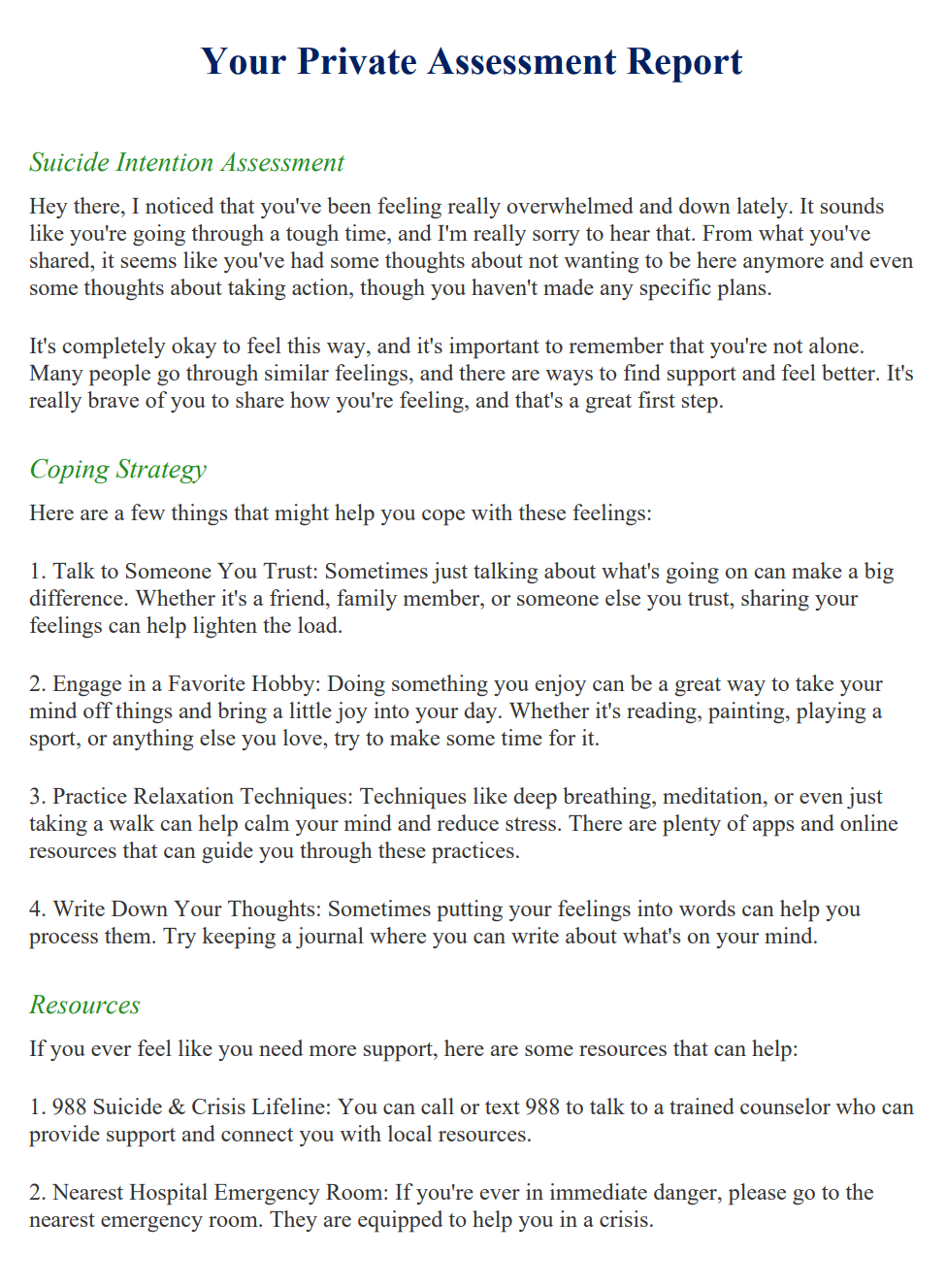}
  \caption{Use Case for Report Generation.}
  \label{fig:Use Case for Report Generation.}
\end{figure}

\subsection{Use Case for General Mental Health QA}
\label{subsec:Appdendix A.3}

The SouLLMate, as illustrated in Figure \ref{fig:Use Case for General Mental Health QA.}, excels in empathy and active listening, offering practical stress relief advice while maintaining professional boundaries. Its responses are well-structured and personalized, using supportive language like "You're not alone."

\end{document}